\documentstyle[prl,aps,epsfig,floats,twocolumn]{revtex} 
\def\ni{\noindent}
\def\sig{{\sigma}}
\def\br{\vec{r}}

\def\bR{\vec{R}}

\def\bff{\vec{f}}

\def\brho{\vec{\rho}}

\def\hC{{\hat C}}

\def\hP{{\hat P}}
\def\hG{{\hat G}}

\def\hI{{\hat I}}

\def\heps{{\hat \epsilon}}

\def\cW{{\cal W}}
\def\cP{\Theta}

\def\lX{{\lambda X}}

\begin{document} 
\draft 
 
 
\title{Granular entropy: Explicit calculations for planar assemblies}
\author{Raphael Blumenfeld and Sam F. Edwards}
 
\address{Polymers and Colloids, Cavendish Laboratory, Madingley Road,  
Cambridge CB3 0HE, UK} 
\maketitle 
\date{\today} 
\maketitle

\begin{abstract} 
This paper proposes a new volume function for calculation of the entropy of planar granular assemblies. This function is extracted from the antisymmetric part of a new geometric tensor and is rigorously additive when summed over grains. It leads to the identification of a conveniently small phase space. The utility of the volume function is demonstrated on several case studies, for which we calculate explicitly the mean volume and the volume fluctuations.
 
\end{abstract} 
\narrowtext 

\vspace{0.5cm}

It has been shown experimentally and computationally \cite{NKBJN}\cite{Makse} that jammed granular systems can be described by statistical mechanics in appropriate circumstances, validating theoretical concepts \cite{EdOa}\cite{MEd}. The simplest approach \cite{EdOa} involves the introduction of compactivity, $X = \partial V/\partial S$ which plays the role of temperature in thermal systems. To quantify $X$ one needs to calculate the entropy $S$ as a function of the volume $V$ and therefore the volume as a function of the position and coordination of the $N$ grains, i.e., a function $\cW$ to complete the analogy between thermodynamics of equilibrium and these non-equilibrium athermal systems
$E\to V \ \ ; \ \ H\to \cW \ \ ; \ \ S(E,V,N)\to S(V,N)$.
This Letter follows a recent analysis \cite {BallBlumenfeld} of planar assemblies in terms of loops and voids. 
Each grain, $g$ can be characterised in terms of the $Z_g$ grains which it is in contact with, $g'$, and the position of these contact points, $\br_{gg'}$ (see figure 1). For each grain we define a centre, $\br_g = (1/Z_g) \sum_{g'=1}^{Z_g} \br_{gg'}$, and vectors $\br_{lg}$ that connect the contact points. The latter vectors form a loop around grain $g$ that is defined to circulate in the clockwise direction. Each vector along this loop can be uniquely identified in terms of the grain $g$ and a neighbouring void $l$. The vectors $r_{lg}$ also form polygons around the voids, whose edges circulate in the anticlockwise direction (see figure 1). To each void we assign a centre $\br_l = (1/Z_l) \sum_{l=1}^{Z_l} \br_{lg}$, where the sum runs over the $Z_l$ grains that surround void $l$. 
Finally, we define a set of vectors $\bR_{lg} = \br_l - \br_g$ that extend from the centre of grain $g$ to the centre of a neighbouring void $l$. This network is self-dual to the $\br$ network so that for each $\bR$ vector there is an $\br$ vector that intersects it. 

We can now define a one-grain geometric tensor $\hC_g$ 

\begin{equation}
C_g^{ij} = \sum_l r_{lg}^i R_{lg}^j 
\label{eq:Bi}
\end{equation}
where the sum runs over all the voids surrounding grain $g$ and $i,j = x,y$ index Cartesian components. Each term in this expression involves only one self-dual pair of vectors and has a straightforward geometrical interpretation:  
Its antisymmetric part is exactly $A_g\heps$, where $\matrix{\heps} = {0 \,1 \choose -1 \,0 }$ and $A_{lg} = {1\over 2} r_{lg} R_{lg} \cos{\alpha_{lg}}$ is the volume of the quadrilateral formed by $\br_{lg}$ and $\bR_{lg}$, shown shaded in figure 1. Its symmetric part measures the deviation of this quadrilateral from a perfect rhombus \cite{BallBlumenfeld}. Note that this holds even when the grain has only two contacts, in which case the quadrilaterals degenerate into triangles. In an isostatic assembly of rough grains $\langle Z_g \rangle = 3$, giving rise to $3N$ quadrilaterals altogether. The volume of the entire system can then be written as a sum over all grains,

\begin{equation}
\cW \hI = \frac{1}{2}\sum_g \hC_g \cdot\heps 
\label{eq:Cii}
\end{equation}
where $\hI$ is the identity matrix. It can also be recast more conveniently as a sum over all the quadrilaterals (henceforth indexed by $n$) that the $\br\!-\!\bR$ pairs make,

\begin{equation}
\cW  = \frac{1}{2}\sum_n r_n R_n \cos{\alpha_n}
\label{eq:Ciii}
\end{equation}
where $r_n=|\br_{lg}|$. This volume function (VF) is additive as required and its simple form makes it convenient for analytical calculations. 

\begin{figure} 
\centerline{\psfig{file=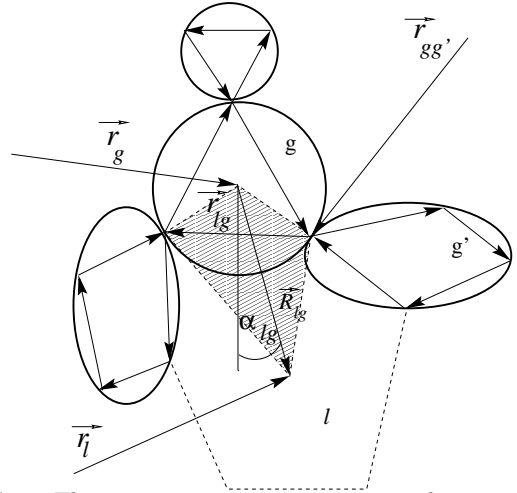,height=6.5cm}} 
\caption{ 
The geometric construction around grain $g$.  The vectors $\br_{lg}$ connect contact points clockwise around each grain $g$ and give rise to anticlockwise loops $l$ around each void. The vectors $\bR_{lg}$ connect from grain centres to loop centres. A one-grain geometric tensor is defined as $\hC_g = \sum_l{\bR_{lg} \br_{lg}}$.}
\end{figure}
 
The partition function can be written as \cite{EdOa}

\begin{equation}
Z = \int e^{-{{\cW(\{q_n\})}\over {\lX}}} \cP(\{q_n\}) \prod_n dq_n
\label{eq:Ci}
\end{equation}
where $\{q_n\}$ is a set of internal degrees of freedom, $\cP(\{q_n\})$ is a probability density function
(PDF) that is subject to the contraint that appropriate grains are in contact. The coefficient $\lambda$ is
the analogue of Boltzmann's constant with $\lX$ ($\equiv 1/\beta$ henceforth) having dimensions of volume ($L^2$ in two dimensions). The analogue of free energy is the effective volume $Y = -\ln Z/\beta$, the (dimensionless) volumetric entropy is $S = \beta^2 \partial Y / \partial \beta$ and the mean volume is $\langle V\rangle = Y + S / \beta$. 

Substituting the VF (\ref{eq:Ciii}) into eq. (\ref{eq:Ci}) we obtain for the partition function

\begin{eqnarray}
Z & = & \prod_n \left[ \int_0^\infty dr_n \int_0^\infty dR_n \int_{-\pi/2}^{\pi/2} d\alpha_n \right] \times \nonumber \\
& \times & e^{-\beta\sum_n r_nR_n\cos{\alpha_n}/2} \cP(\{r_n\}, \{R_n\}, \{\alpha_n\}) 
\label{eq:Civ}
\end{eqnarray}

The values of the variables $R$, $r$ and $\alpha$ are constrainewd by a minimal volume that the assembly can attain,
$v_{min}$, below which further compactification is impossible without causing grains to overlap, and a
maximal volume, $v_{max}$, above which dilatation is impossible without the loss of mechanical equilibrium and fluidisation of the assembly. 
Although these variables are correlated (see below) it is instructive to first assume independent distributions.

\vspace{0.3cm}
\ni {\bf A. Correlation-free systems}:
When the degrees of freedom are independent, $\cP$ can be written as a product of probability density functions (PDFs) of the individual variables. As a first example, let us consider the following: 

\begin{equation}
\cP = \prod_{n=1}^{\langle Z_g\rangle N} 
\frac{\delta(R_n-R_0)\delta(\gamma_n-\gamma_0)}{r_{max}-r_{min}} \ \ ;\ \ r_{min}<r<r_{max}
\label{eq:Di}
\end{equation}
Although simplified, this form yields the essential behaviour observed in experiments \cite{NKBJN}\cite{Makse}. Substituting (\ref{eq:Di}) in (\ref{eq:Civ}), the one-grain partition function is found to be

\begin{equation}
z \equiv Z^{1/N} = \left[\frac{e^{-\beta v_{min}}}{\beta \Delta v}\left( 1 - e^{-\beta \Delta v}\right)\right]^{Z_g}
\label{eq:Dii}
\end{equation}
where $Z_g$ is the coordination number of the grain and the volumes $v_{min}$, $v_{max}$, and $\Delta v$ are, respectively, $R_0\gamma_0 r_{min}/2$, $R_0\gamma_0 r_{max}/2$, and $R_0\gamma_0 (r_{max}-r_{min})/2$. The volumetric entropy can be directly calculated from this expression and the mean volume is 

\begin{equation}
\langle v \rangle = Z_g\left[\frac{v_{min}+v_{max}}{2} + \frac{1}{\beta} - \frac{\Delta v}{2} {\rm ctgh} \left(\frac{\beta \Delta v}{4}\right) \right]
\label{eq:Diii}
\end{equation}
and the volume fluctuations are

\begin{equation}
\langle \delta v^2 \rangle = Z_g \left[ \frac{1}{\beta^2} - \left(\frac{\Delta V}{{\rm sinh}(\beta\Delta V)}\right)^2\right]
\label{eq:Diiia}
\end{equation}
When $\beta\to\infty (\lX\to 0)$ we find that $\langle v \rangle \to Z_g v_{min}$ and $\langle \delta v^2 \rangle \to 0$, while when $\beta\to 0 (\lX\to \infty)$ $\langle v \rangle \to Z_g \left[(v_{min}+v_{max})/2 - \beta\Delta v^2/3\right]$ and $\langle \delta v^2 \rangle \to \Delta v^2/3$. It is interesting to note that expression (\ref{eq:Diii}) has the same form as the one-dimensional result \cite{EdOa}\cite{Models}. The current formalism extends that result to two dimensions and makes it possible to associate it with a particular distribution of degrees of freedom. 

Let us consider a more realistic case

\begin{equation}
\cP = \prod_{n=1}^{\langle Z_g\rangle N} P(r_n) \delta(R_n-R_0) C_{\gamma} e^{-\frac{(\gamma_n-\gamma_0)^2}{2\sig_\gamma^2}}
\label{eq:Div}
\end{equation}
where $\sig_\gamma\ll 1$, $C_{\gamma}$ is the normalisation constant of the PDF of $\gamma$, and $P(r_n)$ is kept arbitrary for the moment. The Gaussian PDF for $\gamma$ around $\gamma_0$, whose value is close to 1, represents a narrow distribution of the angle $\alpha_{n}$ around zero and therefore small devations from a rhombus \cite{corollary}. As we argue below, $P(R_n)$ is Gaussian-like and is narrower than $P(r_n)$, which justifies its approximation as a delta-function. Integration over all $\gamma_n$ and $R_n$ gives

\begin{equation}
Z = \left[ \int dr P(r) \psi(r) e^{\frac{( \beta\sig_\gamma v)^2}{2\gamma_0^2}-\beta v} \right]^{\langle Z_g\rangle N} 
\label{eq:Dv}
\end{equation}
where 
$$\psi(r) = \left[\phi\left(\frac{u}{\sqrt{2}} + \frac{1-\gamma_0}{\sqrt{2}\sig_\gamma}\right) - \phi\left(\frac{u}{\sqrt{2}} + \frac{\gamma_{min}-\gamma_0}{\sqrt{2}\sig_\gamma}\right)\right] \frac{\sig_\gamma C_\gamma}{\sqrt{2/\pi}}$$ 
$\phi$ is the error function and $u = (\sig_\gamma / \gamma_0) \beta v$. For $u\sig_\gamma \ll 1$ (high compactivity)

\begin{equation}
Z \approx \left[ \int_{r_{min}}^{r_{max}} P(r) e^{-\beta \chi v} dr \right]^{Z_g}  
\label{eq:Dvi}
\end{equation}
where 
$$\chi = 1 - \left[ e^{-(1-\gamma_0)^2/(2\sig_\gamma^2)} - e^{-(\gamma_{min}-\gamma_0)^2/(2\sig_\gamma^2)}\right] \frac{C_\gamma \sig_\gamma^2}{\gamma_0} $$
For example, assuming that $P(r)$ is uniform gives that the mean volume per grain is 

\begin{equation}
\langle V \rangle \approx Z_g \chi \left[ \frac{v_{max} + v_{min}}{2} + \frac{1}{\beta\chi} - \Delta v\ {\rm ctgh} (\beta\chi\Delta v) \right]
\label{eq:Dvii}
\end{equation}
and the volume fluctuations are

\begin{equation}
\langle \delta v^2 \rangle = Z_g \left[ \frac{1}{\beta^2} - \left(\frac{\chi\Delta V}{{\rm
sinh}(\chi\beta\Delta V)}\right)^2\right]
\label{eq:Dviia}
\end{equation}
The analysis of these expressions as $\beta\to 0,\infty$ is straightforward and gives the same results as for the previous case with $\Delta v$ replaced by $\chi\Delta v$. Similarly, it is straightforwartd to evaluate these quantities for low compactivity.

The expressions for the volume fluctuations can be now related to response functions and diffusion processes in granular systems. It is unclear at this stage how large is the error introduced by the assumption of independence of the variables. One advantage of the VF proposed here is that it makes it possible to elucidate this issue.

\vspace{0.3cm}
\ni {\bf B. Correlated systems}:
To go beyond the assumption of independence, the key observation is that correlations arise from two sources: (1) relations between the vectors $\br$ and (2) the self-duality of the $r$- and $R$-networks. 
The former can be traced to loops. Whenever $M$ $\br_{lg}$ vectors close a loop their sum vanishes and one of them (2 degrees of freedom) can be expressed in terms of the other $M-1$ vectors. This introduces two $\delta$-functions into the Jacobian, reducing the phase space dimensionality by two. The smallest such loops occur around the grains, $\sum_{l=1}^{Z_g}\br_{lg}=0$. When these are independent each such loop contributed to the Jacobian a term of the form $\prod_{i=1}^{Z_g} P(\br_i)$. The correlations force a modification of this term to [$\prod_{i=1}^{Z_g-1} P(\br_i)] \delta(\br_n + \sum_{i=1}^{Z_g-1}\br_i)$. Note that the value of $Z_g$ is distributed throughout the system, giving rise to a corresponding distribution of such terms in the partition function. 
Voids are also surrounded by loops of $\br$ vectors (e.g., void $l$ in figure 1), each of which further reducing the
phase space dimensionality by two. Grain and void loops are the only irreducible loops in the system (namely, all other loops can be decomposed into these) and therefore only they give rise to correlations of the first type. This observation has an interesting implication: An isostatic systems of $N$ grains has $3N$ $\br$-vectors. The $N$ granular loops and the $N/2$ void loops yield $3N/2$ of these dependent on the rest. Namely, only {\it half} the $\br$ vectors are independent, giving $3N$ degrees of freedom. 

Turning to the second type of correlations, recall that $\bR_{lg}$ extends from the centroid of grain $g$, $\frac{1}{Z_g}\sum_{l'=1}^{Z_g-1}(Z_g-l')\br_{l'g}$, to that of loop $l$, $\frac{1}{Z_l}\sum_{g'=1}^{Z_l-1}(Z_l-g')\br_{lg'}$. Therefore, it can be expressed as a {\it linear combination} of the vectors forming the $g$ and $l$ loops, $\bR_{lg} = \sum_{k=1}^{Z_l + Z_g - 2} a_n \br_n$,
and is uniquely defined in terms of the $\br$-network. On average, an $\bR$ vector depends on $\langle Z_g + Z_l - 2\rangle = 7$ $\br$ vectors and, since there are two grains to a void, then $\langle R\rangle \approx \sqrt{2} \langle r\rangle$. It follows that $P(R)$ can be safely approximated by a Gaussian around this value. 
The dependence of the $R$ variables on the $r$'s further means that the $\gamma$ variables can also be expressed in terms of these because $\gamma_n = \sqrt{1 - \bR_n\cdot\br_n}$. In fact, $\gamma_n$ depends on average on eight $r$ variables and therefore $P(\gamma_n)$ can also be approximated as a Gaussian.
Combining all the above, the VF can be written as

\begin{equation}
\cW  = \frac{1}{2}\sum_{n,m=1}^{3N/2} a_{nm} r_n^x r_m^y 
\label{eq:Fiiib}
\end{equation}
The coefficients $ 0 < a_{nm} < 1$ are rational and form a sparse matrix of zeroes on the diagonal and, on average, seven finite elements in each row. Although appealing, this quadratic form is only useful with particular forms of $P(\{ r \})$ and even then it requires knowledge of the statistics of $a_{nm}$. Its main disadvantage is in mixing the basic quadrilateral volume units and so losing the clear geometrical interpretation of eq. (\ref{eq:Ciii}). 

Although the correlations reduce the phase space dimensionality to $3N$, it is difficult to take them all explicitly into consideration. To make progress, we consider all the original $9N$ variables but include the lowest order correlations - those coming from the intragranular loops, which are the smallest. The justification for this approximation is that since the $R$ and $\gamma$ variables depend on several $r$ degrees of freedom then they are more narrowly distributed and so can be considered as background fluctuations. We believe that this approximation captures the correct physics in granular assemblies.

To illustrate the effects of the intragranular loops, consider a triply coordinated circular grain of diameter $D$. The three contacts on the circumference form a triangle of vectors $\br_{lg}$ and we wish to know the probability $P(r_1,r_2,r_3)$ that the lengths of its sides lie inside $(r_1, r_1+dr_1)$, $(r_2, r_2+dr_2)$, and $(r_3, r_3+dr_3)$. The PDF of one of the sides falling between $r$ and $r+dr$ is

\begin{equation}
P(x=r/D) = 1/\left(\pi \sqrt{ 1 - x^2}\right)
\label{eq:Fii}
\end{equation}
The first two sides can be chosen independently, but once these are in place, the third side is determined by 

\begin{eqnarray}
x_3^2 \equiv f^2(x_1,x_2) = & 2 & x_1x_2\sqrt{(1-x_1^2)(1-x_2^2)} \nonumber \\ 
& + & (x_1^2+x_2^2) - 2x_1^2x_2^2 
\label{eq:Fiii}
\end{eqnarray}
and therefore 

\begin{equation}
P(x_1,x_2,x_3) = {{\delta\left( x_3 - f(x_1,x_2)\right) } \over { \pi^2 \sqrt{\left( 1 - x_1^2 \right) \left( 1 - x_2^2\right) }}} 
\label{eq:Fiiia}
\end{equation}
Although this analysis can be extended to $Z_g > 3$ and to non-circular grains, we do not give it here. We next demonstrate the effect of intragranular correlations on the effective volume.
 
Consider a large ensemble of $N$ randomly assembled monodisperse circular grains, each contacting exactly three neighbours. Ignoring void loop correlations makes the grains effectively independent but not the quadrilaterls. We take a Gaussian form for $P(\gamma)$ and practically a delta-function for $R$. The one-grain partition function ($z = Z^{1/N}$) is then 

\begin{eqnarray}
z = \int \prod_{n=1}^3 & \left[ e^{-\beta r_n R_n \gamma_n / 2} P(R_n) P(\gamma_n)dr_n dR_n d\gamma_n \right] \times \nonumber \\
& P(r_1)P(r_2) \delta\left(r_3 - f(r_1,r_2)\right)
\label{eq:Fiv}
\end{eqnarray}
where the lengths $r_n$ are measured in units of the grain diameter $D$. Integrating over the $R$ and $\gamma$ variables yields

\begin{eqnarray}
z = \int & dr_1 dr_2 P(r_1) P(r_2) \psi(r_1) \psi(r_2) \psi(f(r_1,r_2)) \times \nonumber \\
& \exp{\left[ \frac{\beta^2 R_0^2\sig_\gamma^2(r_1^2 + r_2^2 + f^2))}{8} -  \frac{\beta R_0 \gamma_0 (r_1 + r_2 + f)}{2}\right]}
\label{eq:Fv}
\end{eqnarray}
Substituting from (\ref{eq:Fiiia}), we can now compute $Y$ and from it the mean volume and its fluctuations. The mean volume is plotted in figure 2 as a function of $\lX$.
The asymptotic value as $X\to\infty$ is the average of the close and loose random packings, while the value at $X\to 0$ is the close random packing volume. In terms of grain number density, these are 0.473 and 0.537, respectively.  

\begin{figure}
\centerline{\psfig{file=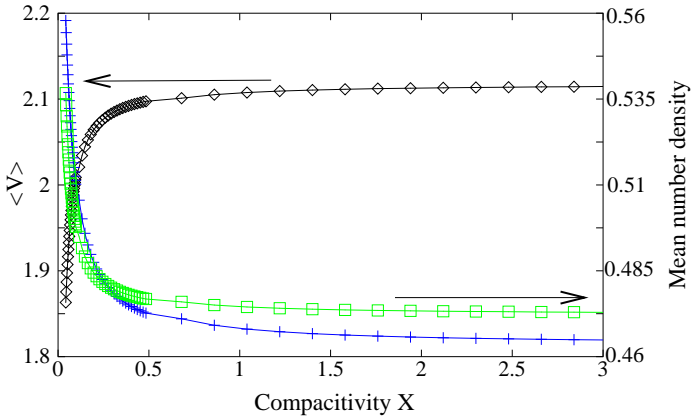,height=5.5cm}}
\caption{
The volume (left axis) and density (right axis) of a trivalent granular assembly of monodisperse circular grains. The moduli of the intragranular vectors connecting the contact points are taken to be correlated in the calculation. The PDFs of $\gamma$ and $R$ are, respectively, Gaussian and a $\delta$-function at the periodic lattice value $R_0=2D/\sqrt{3}$. Two density plots are presented: one assuming independent $r$ variables (squares) and the other taking the intragranular correlations into account (+). For $X > X_0 = 0.265$ the correlations compactify the system while below this value they lead to an opposite effect for the reason discussed in the text. A significant feature is that the difference between the two plots is quite small. The densities of the two calculations differ only by 0.009 ($\approx 2\%$) at $\lX = 5$ and by $0.021$ at $\lX = 0.04$ ($\approx 4\%$).
}
\end{figure}

We can now assess the assumption of independence. Using form (\ref{eq:Fiiia}) for $P(r)$ of all three quadrilaterlas, we also plot in figure 2 the density of the system with and without correlations. It can be observed that the correlations reduce the density of the system at high compactivity, but increase it compared to the correlation-free system at low compactivity. The crossover is at $X_0 = 0.265$. This is expected because the sides cannot all assume simultaneously either $r_{max}$ or $r_{min}$. This constrains the high compactivity limit from overexpanding and the low compactivity regime from overshrinking. In spite of this fundamental difference, it is rather suprising how close the densities of the two systems are over the entire range. The densities of the correlated and uncorrelated systems in the high compactivity limit are, respectively, $0.473$ and $0.464$ ($1.9\%$ difference), while at the low compactivity limit ($\lX=0.04$) the values are $0.537$ and $0.558$ ($3.8\%$ difference). This suggests that simple models that do not take the correlations into account may prove sufficiently accurate for various purposes.

To conclude, we have proposed a new VF for planar granular assemblies that is exactly additive over grains. Nevertheless, we have identified the quadrilaterals, rather than the grains, as the basic units of the system because it is their volumes that cover the entire volume. A sum over grain volumes without involving the quadrilaterals has the disadvantage of requiring an assumption on approximate grain volume. We point out that the VF proposed here applies to any assembly in mechanical equilibrium, not only to isostatic ones. We have found that the relevant phase space is of relatively small dimensionality and identified the sources of geometrical correlations in the system. We have calculated the effective volume and volume fluctuations of several model assemblies both with and without correlations. The identification of quadrilaterlas as the basic building blocks is relevant to modelling. For example, a two-volume model for grains, $v_1$ and $v_2$ with occurrence probabilities $p_1$ and $p_2$, gives a mean volume of \cite{Models}
  
\begin{equation}
\langle V_g \rangle = N\left[ \frac{v_1 + v_2}{2} + \Delta v{\rm tanh}\left(\beta\Delta v + \ln{\sqrt{\frac{p_1}{p_2}}}\right) \right]
\label{eq:Fvi}
\end{equation}
Applying the same model to quadrilaterals gives a similar expression, but it corresponds to a wider distribution of grain volumes both because a grain has $Z_g$ quadrilaterals and because $Z_g$ itself is distributed. Thus, quadrilaterals as basic units yield more realistic models without investment of extra effort.

\vspace{0.5in}

\ni {\bf Acknowledgements} 
 
\ni We acknowledge discussions with Prof R. C. Ball and Dr D. V. Grinev.


\begin{thebibliography}{99}
\bibitem{corollary} This is related to the result in \cite{BallBlumenfeld} that the volume average of the symmetric part of $\sum_g \hC_g$ vanishes as $1/\sqrt{N}$.
\end{thebibliography}
\end{document}